\documentclass[aps,prl,twocolumn,showpacs,floatfix,superscriptaddress]{revtex4}
\usepackage{graphicx,amsmath,amssymb,ulem
}
\usepackage{color}

\newcommand{\be}{\begin{equation}}
\newcommand{\ee}{\end{equation}}
\newcommand{\bea}{\vspace{0.25cm}\begin{eqnarray}}
\newcommand{\eea}{\end{eqnarray}}

\begin{document}
\title{ Experimental realisation of quantum illumination}

\author{E.~D.~Lopaeva}
\affiliation{INRIM, Strada delle Cacce 91, I-10135 Torino, Italy}
\affiliation{Dipartimento di Fisica, Politecnico di Torino, I-10129, Torino, Italy}

\author{I. Ruo Berchera}
\affiliation{INRIM, Strada delle Cacce 91, I-10135 Torino, Italy}

\author{I.~P.~Degiovanni}
\affiliation{INRIM, Strada delle Cacce 91, I-10135 Torino, Italy}

\author{S.~Olivares}\affiliation{Dipartimento
di Fisica, Universit\`a degli Studi di Milano, I-20133 Milano,
Italy}
\affiliation{Dipartimento di Fisica, Universit\`a degli Studi
di Trieste, I-34151 Trieste, Italy} \affiliation{CNISM UdR Milano
Statale, I-20133 Milano, Italy}

\author{G.~Brida}
\affiliation{INRIM, Strada delle Cacce 91, I-10135 Torino, Italy}

\author{M.~Genovese}
\affiliation{INRIM, Strada delle Cacce 91, I-10135 Torino, Italy}

\begin{abstract}
We present the first experimental realisation of the quantum illumination protocol proposed in
Ref.s [S. Lloyd, Science {\bf 321}, 1463 (2008); S. Tan {\it et al.}, Phys. Rev. Lett. {\bf 101},
253601 (2008)], achieved in a simple feasible experimental scheme based on photon-number
correlations. A main achievement of our result is the demonstration of a strong robustness of the
quantum protocol to noise and losses, that challenges some widespread wisdom about quantum
technologies.
\end{abstract}
\maketitle

\maketitle


 Properties of quantum states have disclosed the possibility of realizing tasks beyond classical limits, originating a field collectively christened quantum technologies \cite{4,6,7,8,9,10,11}. Among them, quantum metrology and imaging aim to improve the sensitivity and/or resolution of measurements exploiting non-classical features, in particular non classical correlations \cite{Kolobov,Treps:03,boy:08,bri:10, gio:11}. However, in most of the realistic scenarios, losses and noise are known to nullify the advantage of adopting quantum strategies \cite{Walmsley-PRL-2011}.  Here, we present the first experimental realization of a quantum enhanced scheme \cite{llo:08,tan:08}, designed to target detection in a noisy environment, preserving a strong advantage over the classical counterparts even in presence of large amount of noise and losses. This work, inspired by theoretical ideas elaborated in \cite{llo:08,tan:08,sha:09,gua:09} (see also\cite{ms}), has been implemented exploiting only photon-number correlations in twin beams and, for its accessibility, it can find widespread use. Even more important, it paves the way to real application of quantum technologies by challenging the common believe that they are limited by their fragility to noise and losses.

Our scheme for target detection is inspired by the "Quantum Illumination" (QI) idea
\cite{llo:08,tan:08}, where the correlation between two beams of a bipartite non-classical state of
light is used to detect the target hidden in a noisy thermal background, which is partially
reflecting one of the beam. In \cite{tan:08,sha:09} it was shown that for QI realized by twin
beams, like the ones produced by parametric down conversion, there exists in principle an optimal
reception strategy offering a significant performance gain respect to any classical strategy.
Unfortunately, this quantum optimal receiver, is not yet devised, and even the theoretical proposal
of sub-optimal quantum receiver \cite{sha:PRA:09} was very challenging from an experimental point
of view, and has not been realized yet.
\par
Our aim is to lead the QI idea to an experimental demonstration in a realistic scenario. Therefore,
in our realization we consider realistic a-priori unknown background, and a reception strategy
based on photon-counting detection and second-order correlation measurements. We demonstrate that
the quantum protocol performs astonishingly better than its classical counterpart based on
classically-correlated light at any background noise level. More in detail, we compare quantum
illumination, specifically twin beams (TWB), with classical illumination (CI) based on correlated
thermal beams (THB), that turns out to be the best possible classical strategy in this detection
framework.
\par
On the one hand our approach, based on a specific and affordable detection strategy in the context
of the current technology, can not aim to achieve the optimal target-detection bounds of Ref.
\cite{tan:08}, based on quantum Chernoff bound \cite{aud:07,cal:08,pir}. On the other hand, it
maintains most of the appealing features of the original idea, like a huge quantum enhancement and
a robustness against noise, paving the way to future practical application because of the
accessible measurement technique. Our study also provides a significant example of ancilla-assisted
quantum protocol, besides the few previous realizations, e.g. \cite{bri:10,bri:ea,tak,al}.
\par

\begin{figure}[tbp]
\begin{center}
 \includegraphics[width=0.5\textwidth]{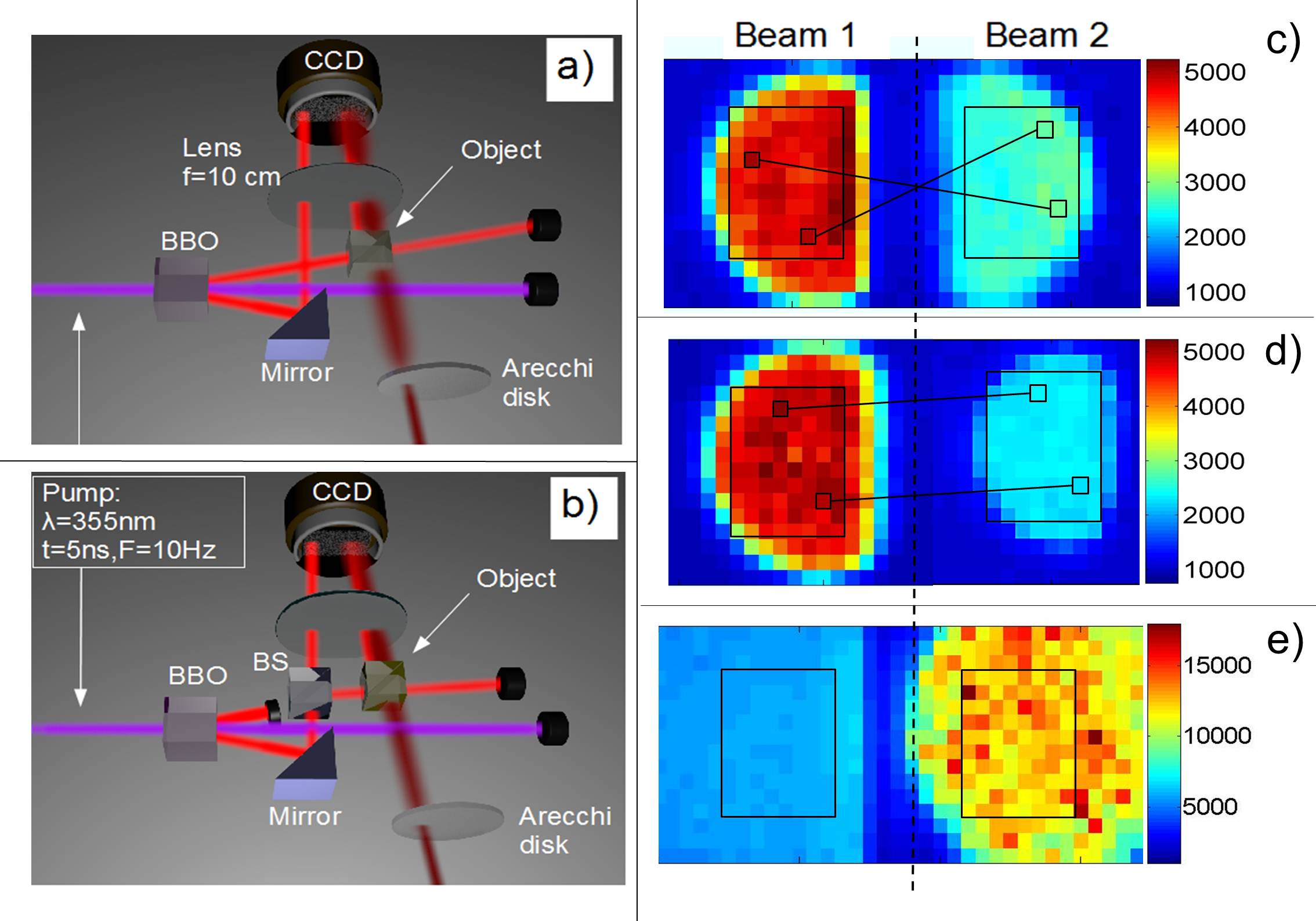}
\caption{Experimental setup. a) Quantum illumination: after the BBO
  crystal, where TWB are generated, one of them (the
  ``ancilla'') is reflected towards the detection system. The
  correlated beam is partially detected, together with the thermal
  field from the Arecchi's disk, when the object (actually a beam
  splitter) is present, otherwise it is lost (not showed). Low-pass
  filter (95~\% of transmission at 710nm) and UV-reflecting mirror,
  not showed, are used to minimize the background noise while
  maintaining low losses. The lens, placed at the focal length from
  the crystal and the CCD camera, realizes the Fourier transform of
  the field at the output face of the crystal. b) Classical
  illumination: one beam from PDC is stopped and the other one is
  split at a beam splitter (BS) for generating correlated
  multi-thermal beams. The power of the pump is adjusted
  to obtain the same energy of the TWB. c) Detected TWB,
  in the presence of the object, without thermal background. The region of
  interest is selected by an interference filter centered around the
  degeneracy wavelength (710~nm) and bandwidth of 10~nm. After
  selection the filter is removed.  d) Detected field for split
  thermal beams in the presence of the object, without thermal
  noise. e) A typical frame used for the measurement, where  a strong thermal background has
  been added on the object branch. The colour scales on the right
  correspond to the number of photons per pixel.
} \label{setup} \end{center} \end{figure}
In our set up (see Fig.~\ref{setup}) Parametric Down Conversion (PDC) is exploited to generate two
correlated light emissions with average number of PDC photons per spatio-temporal mode $\mu=0.075$,
that are then addressed to a high quantum efficiency CCD camera (See Supplementary Information,
Sec.I). In the QI protocol (Fig.~\ref{setup}a) one beam is directly detected, while a target object
(a 50:50 beam splitter) is posed on the path of the other one, where it is superimposed with a
thermal background produced by scattering a laser beam on an Arecchi's rotating ground glass. When
the object is removed, only the background reaches the detector.  The CCD camera detects, on
different areas, both the optical paths. In the CI protocol (Fig.~\ref{setup}b), the TWB are
substituted with classical correlated beams, obtained by splitting a single arm of PDC, that is a
multi-thermal beam, and by adjusting the pump intensity to ensure the same intensity, time and
spatial coherence properties for the quantum and the classical sources.
\par

We measured the correlation in the photon number $N_{1}$ and $N_{2}$ detected by pairs of pixels
intercepting correlated modes of beam "1" and "2" respectively (Fig.~\ref{setup} c-d-e),
\cite{Brambilla-PRA-2004,pra}. With our experimental setup, this correlation can be evaluated with
a single image by averaging over all the $N_{\rm pix}$ pixels pairs. Albeit the usage of spatial
statistic is not strictly necessary, it is practically effective and allows to reduce the
measurement time (less images needed)\cite{Bri-OptExp:10}.
\par
\begin{figure}[tbp]
\begin{center}
 \includegraphics[width=0.5\textwidth]{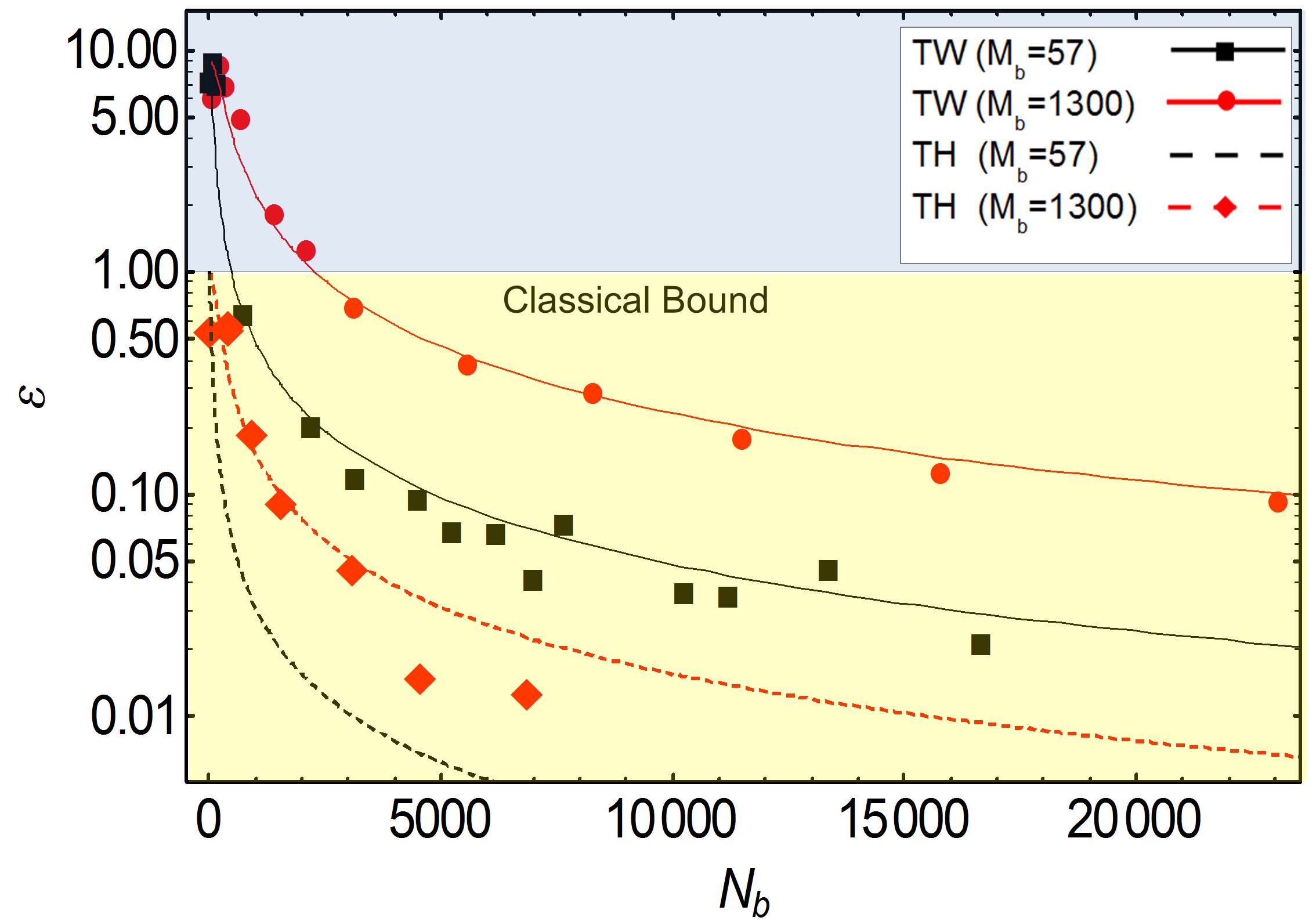}
\caption{Generalized Cauchy-Schwarz parameter $\varepsilon$  in the case of twin beams, $\varepsilon^{(\rm TW)}$, and of the correlated thermal beams, $\varepsilon^{(\rm TH)}$, as a function of the average number
  of background photons $N_{b}$ for a number of background modes $M_{b}=57$ (black series) and
  $M_{b}=1300$ (red). The lines represent the theoretical prediction at $\mu=0.075$ (the last estimated independently).
  } \label{CSfig} \end{center}
\end{figure}

In order to quantify the quantum resources exploited by our QI strategy we introduce a suitable
non-classicality parameter: the generalized Cauchy-Schwarz parameter $\varepsilon = \langle :
\delta N_{1} \delta N_{2} : \rangle ~ / ~ \sqrt{\langle :\delta^{2} N_{1}:\rangle\langle
:\delta^{2} N_{2}:\rangle}$, where $\langle:~:\rangle$ is the normally ordered quantum expectation
value and $\delta^2 N_i = (N_i - \langle N_i \rangle)^2$ the fluctuation of the photon number
$N_{i}$, $i=1,2$.  This parameter is interesting since it does not depends on the losses and it
quantifies non-classicality: $\varepsilon \leq 1$ for classical state of light (with positive
Glauber-Sudarshan $P$-function), while quantum state with negative/singular $P$-function can
violate this bound \cite{Sekatski2012}.  In Fig.~\ref{CSfig} we report the measured $\varepsilon$
and the theoretical prediction. One observes that for TWB $\varepsilon^{(QI)}$ is actually in the
quantum regime ($\varepsilon^{(QI)} >1$) for small values of the thermal background $\langle N_b
\rangle $; in absence of it ($\langle N_b \rangle =0$) we obtain $\varepsilon^{(QI)}_{0}\simeq 10$.
As soon as the contribution of the background to the fluctuation of $N_{2}$ becomes dominant,
$\varepsilon^{(QI)}$ decreases quite fast, well below the classical threshold.  As expected, for
THB $\varepsilon^{(CI)}$ is always in the classical regime, and it is equal to one for $\langle N_b
\rangle =0$.
\par

We consider an a-priori unknown background, meaning that it is impossible to establish a  reference
threshold of photo-counts (usually the mean value of the background) to be compared with the
possible additional mean photo-counts coming from the reflected probe beam (if the target is
present). Therefore, the estimation of the first order (mean values) of the photo-counts
distribution, typical of other protocols (e.g. \cite{Treps:03, bri:10, gio:11}), is here not
informative regarding the presence/absence of the object. We underline that this unknown-background
hypothesis accounts for a ``realistic'' scenario where background properties can randomly change
and drift with time and space.
\par

\begin{figure}[tbp]
\begin{center}
 \includegraphics[width=0.5\textwidth]{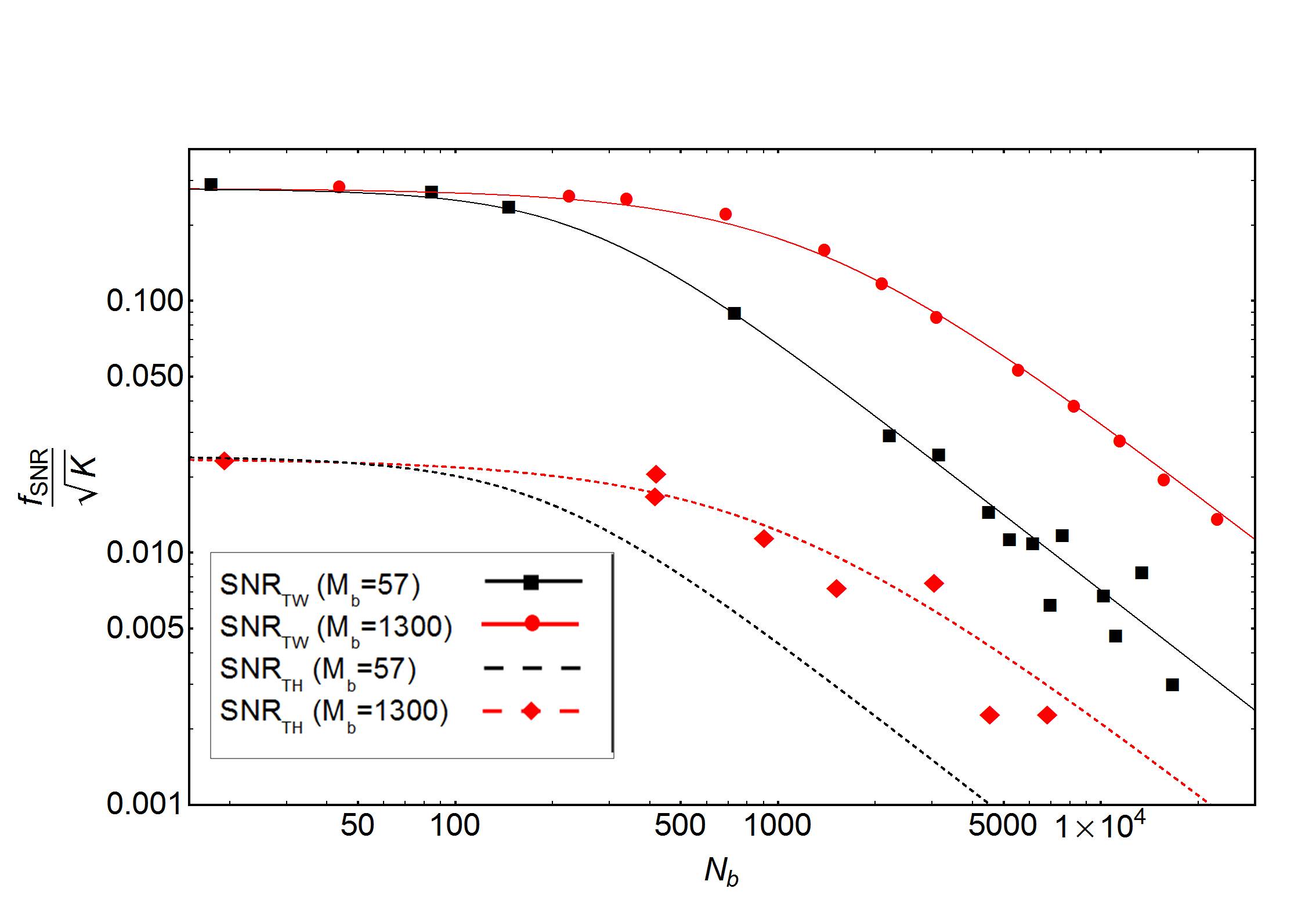}
\caption{Signal-to-noise ratio (SNR) versus the number of background photons $\langle N_{b}\rangle$
normalized by the square root of number of
  realization. The red (black) markers refer to
  $M_{b}=1300$ ($M_{b}=57$) and the solid (dashed) theoretical curve
  corresponds to quantum (classical) illuminating beams. The estimation of quantum mean values of
Eq.~(\ref{SNR}) is obtained by performing averages of $\Delta_{1,2}^{\rm (in/out)}$ over a set of $N_{\rm img}$ acquired images
($N_{\rm img}=2000,4000,6000$ for twin beams at $M_{b}=1300$, $M_{b}=57$ and thermal beams at $M_{b}=1300$, respectively).
The lowest
  curve of the classical protocol has not been compared with the
  experimental data because the SNR is so low that a very large number
  of images (out of the possibility of the actual setup) is required
  to get reliable points.}
\label{SNR} \end{center} \end{figure}

\begin{figure}[tbp] \begin{center}
 \includegraphics[width=0.5\textwidth]{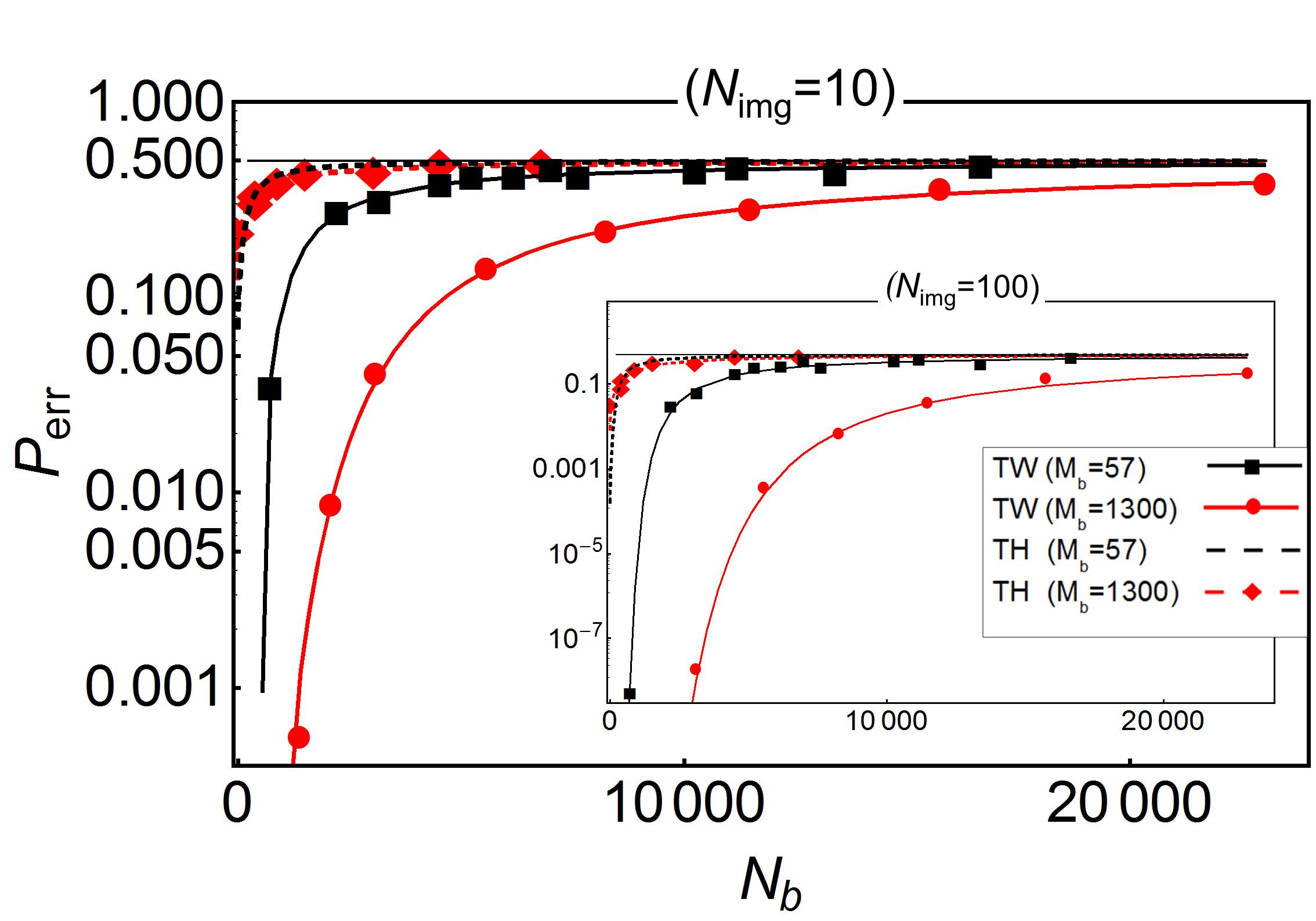}
\caption{Error probability $P_{\rm err}$ of the target detection
  versus the total number of photons of the thermal background $N_{b}$
  evaluated with $N_{\rm img}=10$ ($N_{\rm img}=100$ in the
  inset). The black squares and red circles are the data for QI with
  $M_{b}=57$ and $M_{b}=1300$, respectively, while red diamonds
  referes to the data for the CI with $M_{b}=1300$. The curves are the
  corresponding theoretical predictions.} \label{Perr} \end{center}
\end{figure}

For this reason we propose to discriminate the presence/absence of the object by distinguishing between the two
corresponding values of the covariance $\Delta_{1,2}$, evaluated experimentally as
\begin{equation} \label{Cov}
\Delta_{1,2}=E[N_{1}N_{2}]-E[N_{1}]E[N_{2}].
\end{equation}
$E[X]= \frac{1}{\mathcal{K}}\sum^{\mathcal{K}}_{k=1} X^{(k)}$ represents the
average over the set of $\mathcal{K}$ realizations that in our
experiment correspond to the number $\mathcal{K}=N_{\rm pix}$ correlated pixels pairs. The signal-to-noise ratio  can
be defined as the ratio of the mean ``contrast'' to its standard
deviation (mean fluctuation):
\begin{equation} \label{SNRdef}
  f_{\rm SNR}\equiv
\frac{\left|\left\langle \Delta_{1,2}^{\rm (in)}-\Delta_{1,2}^{\rm (out)}\right\rangle\right|}
  {\sqrt{\left\langle\delta^{2}\left(\Delta_{1,2}^{\rm (in)}\right)\right\rangle+\left\langle\delta^{2}\left(\Delta_{1,2}^{\rm (out)}\right)\right\rangle}},
\end{equation}
where ``in'' and ``out'' refer to the presence and absence of the object.
\par
\begin{figure}[tbp]
\begin{center}
 \includegraphics[width=0.5\textwidth]{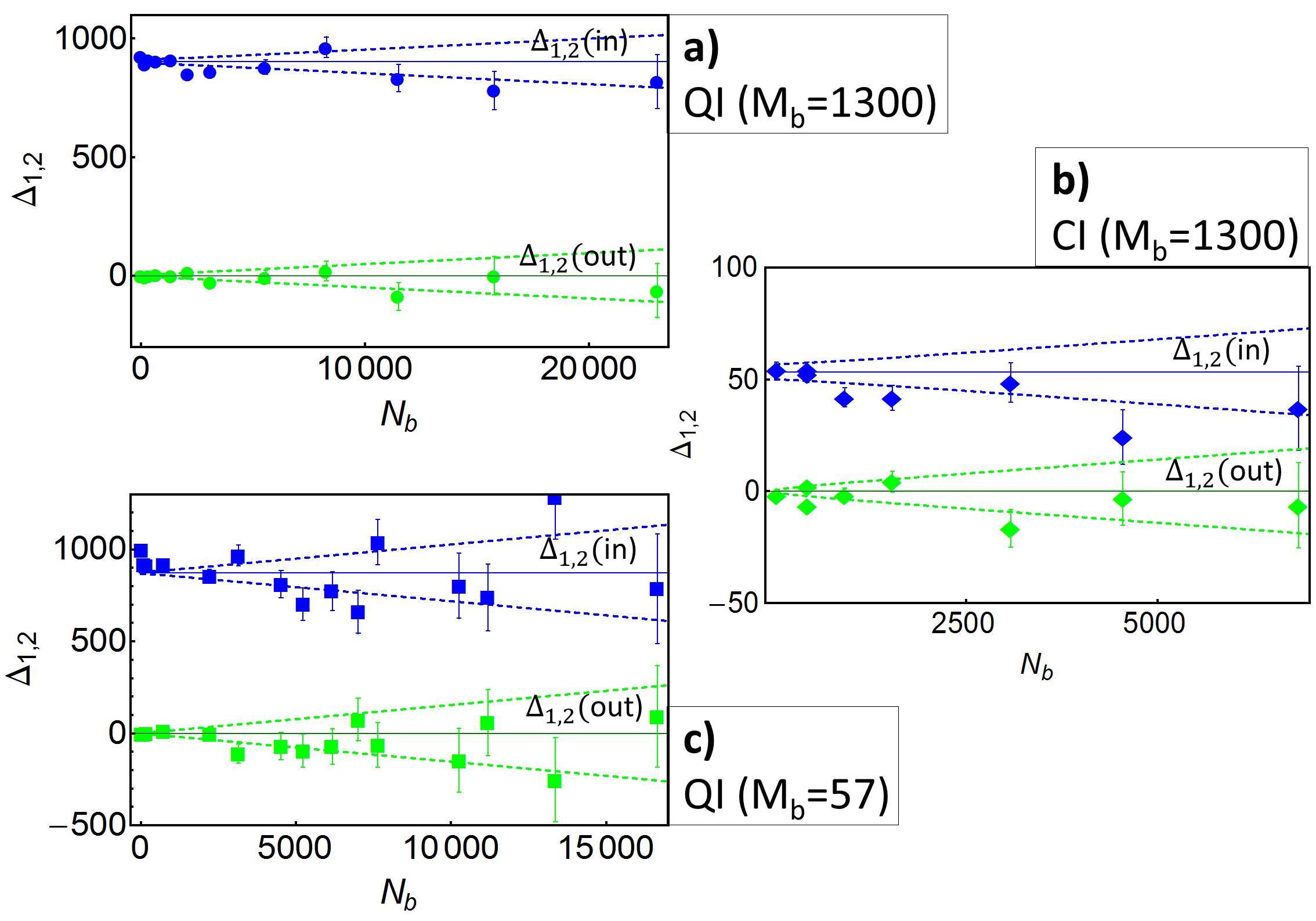}
\caption{Covariance in the presence, $\Delta_{1,2}^{\rm (in)}$ (blue),
  or absence, $\Delta_{1,2}^{\rm (out)}$ (green), of the target. a)
  and b) refer to QI and CI, respectively, for the same number of background
  modes $M_{b}=1300$; c) refers to QI with a a lower number of modes,
  $M_{b}=57$. Uncertainty bars represent the effect of the background noise
  on the covariance estimation (obtained averaging over the $N_{\rm img}=2000,6000$ and
  $4000$ images in the graphs from the top to bottom, respectively). Horizontal lines are the theoretical values
  $\langle\Delta_{1,2}^{\rm (in/out)}\rangle$, while the dashed lines are the uncertainty interval evaluated theoretically as
 $ \langle \delta^{2}  \Delta_{1,2}^{\rm (in/out)}\rangle/\sqrt{N_{\rm img}}$.  }
\label{Cov.fig} \end{center} \end{figure}

For $\mathcal{K}>>1$, the ``contrast'' at the numerator of Eq. (\ref{SNRdef}) corresponds to the
quantum expected value of the covariance i.e. $\langle \Delta_{1,2}^{\rm (in)}\rangle \simeq
\langle \delta N_{1}\delta N_{2}\rangle$, where obviously $\langle \Delta_{1,2}^{\rm
(out)}\rangle=0$. For a generic prominent background with a mean square fluctuation $\langle
\delta^{2}N_{b}\rangle$, the ``noise'' at the denominator depends only on the local statistical
properties of the beam 1 and of the uncorrelated background, i.e.  $\langle
\delta^{2}\Delta_{1,2}\rangle\simeq\langle\delta^{2}N_{1} \rangle \langle \delta^{2}N_{b}\rangle$
(Supplementary Information, Sec.II-a). This  is shown in Fig.~\ref{Cov.fig}, where the estimated
covariance of Eq. (\ref{Cov}) is plotted versus the intensity of the thermal background, used in
our experiment. As expected, the average value of covariance does not depend on the quantity of
environmental noise, while the uncertainty bars do.
\par
While the signal-to-noise ratio unavoidably decreases with the added noise for both QI and CI, the
quantum enhancement parameter $(R= f^{(QI)}_{\rm SNR}/f^{(CI)}_{\rm SNR})$ in the presence of
dominant background and equal local resources becomes
\begin{equation}\label{R}
R\approx\langle \delta N_{1}\delta N_{2}\rangle_{QI}/\langle \delta N_{1}\delta N_{2}\rangle_{CI}.
\end{equation}
Being $R$ expressed as a ratio of covariances, it is remarkably independent on the amount of
losses, noise and reflectivity of the object.
\par
According to Eq. (\ref{R}) the enhancement is lower bounded by the amount of violation of the
Cauchy-Schwarz inequality  for the quantum state considered in the absence of background, i.e.
$R\simeq\varepsilon^{(QI)}/\varepsilon^{(CI)}\geq\varepsilon^{(QI)}_0$. The equality holds for
classical states saturating the classical bound, $\varepsilon^{(CI)}_0 = 1$  (Supplementary
Information,Sec.B).
\par
In particular, in our experiment we  compared the performance of TWB with a classically correlated
state with $\varepsilon^{(CI)}_0 = 1$ (hence representing the best possible classical strategy),
i.e. with a split thermal beams presenting the same local behaviour of the TWB. In this case the
enhancement can be explicitly calculated obtaining $R\simeq(1+\mu)/\mu$, hence the quantum strategy
performs orders of magnitude better than the classical analogous when $\mu \ll 1$, namely when a
low intensity probe is used.
\par
Incidentally, since covariance is always zero (i.e. $\varepsilon = 0$) when using split coherent
beams, they do not provide a valuable alternative in the situation considered here, i.e. when first
order momenta are not informative due to unknown fluctuating background.
\par

In Fig.~\ref{SNR}, the theoretical prediction for $f_{\rm SNR}/\sqrt{\mathcal{K}}$ is compared with
the experimental data. In perfect agreement with theory, the quantum enhancement is almost constant
($R\gtrsim 10$) regardless the value of $ \langle N_{b} \rangle$.  Therefore, the measurement time,
i.e. the number of repetitions $N_{\rm  img}$ needed for discriminating the presence/absence of the
target, is dramatically reduced in QI (for instance, to achieve $f_{\rm SNR}=1$, $N_{\rm img}$ is
almost 100 times smaller when quantum correlations are exploited).
\par
Another figure of merit that highlights the superiority of the quantum strategy versus the
classical one is the the error probability in the discrimination of the presence/absence of the
target ($P_{\mathrm{err}}$).  In Fig.~\ref{Perr} we report $P_{\mathrm{err}}$ versus the number of
photons of the thermal background $\langle N_{b} \rangle$. $P_{\mathrm{err}}$ is estimated fixing
the threshold value of the covariance that minimizes the error probability itself. Fig.~\ref{Perr}
shows a remarkable agreement between the theoretical predictions (lines) and the experimental data
(symbols), both for QI and CI strategy.  $P_{\mathrm{err}}$ of QI is several orders of magnitude
below the CI one and, in terms of background photons, the same value of the error probability is
reached for a value of $N_{b}$ at least 10 times smaller than in the QI case.
\par
In conclusion, we demonstrated experimentally quantum enhancement in detecting a target in a
thermal radiation background. Our system shows quantum correlation ($\varepsilon{(QI)}_0 \simeq
10$) even in presence of losses introduced by a partially reflective target. Remarkably, even after
the transition to the classical regime ($\varepsilon^{(QI)} \ll 1$) due to the presence of
background ($\langle N_{b} \rangle \gg 1$), the scheme preserves the same strong advantage with
respect to the best classical counterpart based on classically correlated thermal beams.
Furthermore, the results are general and do not depend on the specific properties of the background
used in the experiment.
\par
In paradigmatic quantum enhanced schemes, often based on the experimental estimation of the first
momenta of the photon number distribution, such as quantum imaging protocol \cite{bri:10},
detection of small beam displacement \cite{Treps:03} and phase estimation by interferometry
\cite{gio:11}, it is well known that losses and noise rapidly decrease the advantage of using
quantum light \cite{Walmsley-PRL-2011}.  This enforced inside the generic scientific community the
common belief that the advantages of entangled and quantum state are hardly applicable in a real
context, and they will remain limited to experiments in highly controlled laboratories, and/or to
mere academic discussions. Our work breaks this belief showing orders of magnitude improvements
compared to CI protocol, independently on the amount of noise and losses using devices available
nowadays.  In summary, we believe that the photon counting based QI protocol, for its robustness to
noise and losses has a huge potentiality to promote the usage of quantum correlated light in real
environment.

{\bf Acknowledgements}
\par
The research leading to these results has received funding from the EU FP7 under grant agreement n.
308803 (BRISQ2), Fondazione SanPaolo and MIUR (FIRB ``LiCHIS'' - RBFR10YQ3H, Progetto Premiale
``Oltre i limiti classici di misura''). SO acknowledges financial support from the University of
Trieste (FRA 2009). The authors thank M.~G.~A.~Paris for useful discussions.

\section{Additional Information on Experimental set-up}

In our set up a pump laser beam at 355 nm (triplicated Nd-Yag Q-switched laser) with  5 ns pulses
at 10 Hz repetition rate, pumps a type II 7 mm long BBO crystal, cut for producing collinear
degenerate emission.

The traveling wave PDC generates a spatially multimode emission, where each mode
corresponds to the transverse component of a specific wavevector. Pairs of correlated modes,
corresponding to opposite transverse components of the wavevector with respect to the pump
direction, are found in symmetric positions in the far field around the degeneracy wavelength of 710 nm[26-28].
Thus, we choose (Fig.~1 c-d-e) two correlated
regions of interests on the CCD array, a Princeton Instrument Pixis 400 BRW ($1340X400$ array with
pixel size of 20 $\mu$m, high quantum efficiency, more than 80\% and low noise in the read out
process, about $4$ electrons/pixel). The proper sizing of the pixels and the centering of the
2-dimensional array with sub-mode precision, allows maximizing for each pair of pixels the
collection of the correlated photons and at the same time minimizing the possible presence of
uncorrelated ones [27,28]. The photon-number-correlation, even at the quantum level for QI, is
realized independently for each pairs of symmetrical (translated) pixels belonging to the
correlated regions of interests of the TWB (THB). The number of pixel pairs we used is
$N_{pix}=80$ with area $A_{\rm pix}=(480\mu\hbox{m})^{2}$. We underline that spatial statistics for the photon-counting-based QI protocol is
completely unnecessary. Since QI does not aim an image reconstruction, two conventional
photo-counters (i.e. without spatial resolution) will be able to accomplish the evaluation of
correlation by temporal statistics, and consequently the QI protocol as well. In our
proof-of-principle experiment spatial resolving detectors were employed to have a better control on
the statistical properties of the field employed, in order to provide a proper comparison between
the experimental data and the theoretical prediction.


The mean number of photons detected per pixel is $\langle N\rangle\simeq 4200$.
The number of collected spatio-temporal modes is estimated to be
$M=9\cdot 10^{4}$ by fitting a multi-thermal statistics. Thus, taking into account the overall transmission and detection efficiency (about 62\% [27]), the average number of
PDC photons per mode is $\mu=\langle N\rangle/(\eta M)\simeq 0.075$.  We measured separately the size of
the spatial mode, as the FWHM of the correlation function between the
two beams, $A_{\rm corr}=(120\pm 20\mu\hbox{m})^{2}$. Therefore, the number of
spatial modes is about $M_{\rm sp}=A_{\rm pix}/A_{\rm corr}=16\pm 5$ and the number of
temporal modes $M_{\rm t}=M/M_{\rm sp}=(6\pm2)\cdot10^{3}$, the last one being
consistent with the ratio between the pump pulse duration and the
expected PDC coherence time, i.e. 1~ps.

The background field is produced by a pulsed laser scattered by a rotating ground glass disk (Arecchi's disk) and collimating optics. 
The spatial properties at the CCD plane are set to similar value of the PDC emission, while the temporal modes $M_{b}$ are selected by adjusting the pulse duration and the speed of the disk.

\section{SNR in case of preponderant background}

\subsection{Evaluation of $f_{\mathrm{SNR}}$ }

For a large number of samples $\mathcal{K}>>1$, the ``contrast'' at the numerator of Eq. (2) of the
main text corresponds to the quantum expected value of the covariance, i.e. $\langle
\Delta_{1,2}^{\rm (in)}\rangle \simeq \langle \delta N_{1}\delta N_{2}\rangle$, while the mean
square fluctuation of the covariance $\langle \delta^{2}\Delta_{1,2}\rangle$ at the denominator can be calculated as

\begin{equation} \label{noise}
\mathcal{K}\left\langle \delta^{2}\Delta_{1,2}\right\rangle\simeq\left\langle \delta^{2}(\delta N_{1}\delta N_{2})\right\rangle\equiv \left\langle \left(\delta N_{1}\delta N_{2}\right)^{2}\right\rangle-\left\langle\delta N_{1}\delta N_{2}\right\rangle^{2}.
\end{equation}

By replacing $\delta N_{2}\mapsto \delta N^{(in)}_{2}+\delta N_{b}$ where  $N^{(in)}_{2}$ is the number
of detected photons that has been reflected by the target, and $N_{b}$ is the uncorrelated background, the
right hand side of Eq. (\ref{noise}) can be rewritten as

\begin{eqnarray} \label{noise-approx}
\mathcal{K}\langle \delta^{2}\Delta_{1,2}\rangle&\simeq &  \left\langle \left(\delta N_{1}\delta N^{(in)}_{2}+\delta N_{1} \delta N_{b}\right)^{2}\right\rangle-\left\langle\delta N_{1}\delta N^{(in)}_{2}+\delta N_{1}\delta N_{b}\right\rangle^{2}\\\nonumber
&=&\left\langle \left(\delta N_{1}\delta N_{2}\right)^{2}\right\rangle-\left\langle\delta N_{1}\delta N_{2}\right\rangle^{2}+\left\langle\delta^{2} N_{1}\right\rangle\left\langle\delta^{2} N_{b}\right\rangle\\\nonumber
&=&\left\langle \delta^{2}(\delta N_{1}\delta N^{(in)}_{2})\right\rangle+\left\langle\delta^{2} N_{1}\right\rangle\left\langle\delta^{2} N_{b}\right\rangle,
\end{eqnarray}
where we used the statistical independence of $N_{b}$ and the fact that $\langle\delta N_{b}\rangle=0$.
It is clear that in the absence of the target (situation labeled with the superscript "out"), $N^{(in)}_{2}=0$, thus  $\langle
\delta^{2}\Delta^{(out)}_{1,2}\rangle=\left\langle\delta^{2}
N_{1}\right\rangle\left\langle\delta^{2} N_{b}\right\rangle$, since nothing is reflected to the
detector. However, if the the background fluctuations $\left\langle\delta^{2} N_{b}\right\rangle$
is the largest contribution to the noise, also when the target is present (indicated with superscript "in") we can write $\langle
\delta^{2}\Delta^{(in)}_{1,2}\rangle\simeq\left\langle\delta^{2}
N_{1}\right\rangle\left\langle\delta^{2} N_{b}\right\rangle$. Under this assumption representing a
realistic situation of a very noisy environment, the SNR becomes
\begin{equation} \label{SNR}
  f_{\rm SNR}\simeq
\frac{\langle \delta N_{1}\delta N_{2}\rangle}
  {\sqrt{2\left\langle\delta^{2} N_{1}\right\rangle\left\langle\delta^{2} N_{b}\right\rangle}}.
\end{equation}
We underline that Eq. (\ref{SNR}) holds for a dominant background, irrespective of its statistics
(e.g. multi-thermal or Poissonian).

In our experiment we consider background with multi-thermal statistics. For a generic multi-thermal
statistics with number of spatiotemporal modes $M$, mean photon number number per mode $\mu$, the
total number of detected photons is $\left\langle N\right\rangle=M \eta \mu$ and the mean squared
fluctuation is $\left\langle\delta^{2} N\right\rangle= M \eta \mu(1+\eta \mu)=\left\langle N
\right\rangle\left(1+\left\langle N\right\rangle/M\right)$ [see for example: L. Mandel, E. Wolf, {\it Optical Coherence and Quantum Optics} (Cambridge University Press, 1995)], where $\eta$ is the detection
efficiency.

Thus, the amount of noise introduced by the background can be increased by boosting  its total
number of photons $\left\langle N_{b}\right\rangle$ or by varying the number of modes $M_{b}$, as
highlighted from the behaviour of the SNR in Fig.2,3 and 4 of the main text.

Moreover, both TWB and correlated THB present locally the same multi-thermal statistics, but with a
number of spatiotemporal modes $M=9\cdot 10^{4}$ much larger than the one used for the background
beam ($M_{b}=57$ in one case and, $M_{b}=1.3 \cdot10^{3}$ in the other). This contributes to make
the condition of preponderant background effective in our realization, even for a relatively small
value of $N_b$.

However, we point out that all the theoretical curves reported in all the Figures are evaluated by
the exact analytical calculation of the four order (in the number of photons) quantum expectation
values appearing on the right hand side of Eq. (\ref{noise}), even if the whole expressions are far
more complex than the ones obtained with the assumption of preponderant background.

\subsection{Quantum enhancement $R$}

Starting from Eq. (\ref{SNR}) and considering the same local resources for classical and quantum
illumination beams (in particular the same local variance $\left\langle\delta^{2}
N_{i}\right\rangle_{CI}=\left\langle\delta^{2}  N_{i}\right\rangle_{QI}$ ($i=1,2$)) the enhancement
of the quantum protocol can be easily obtained as

\begin{equation}\label{R}
R=\frac{f^{(QI)}_{\rm SNR}}{f^{(CI)}_{\rm SNR}}\approx\frac{\langle\delta N_{1}\delta N_{2}\rangle_{QI} }{\langle \delta N_{1}\delta N_{2}\rangle_{CI}}= \frac{\varepsilon^{(QI)}}{\varepsilon^{(CI)}}.
\end{equation}
with $\varepsilon = \langle : \delta N_{1} \delta N_{2} : \rangle ~ / ~ \sqrt{\langle :\delta^{2}
N_{1}:\rangle\langle :\delta^{2} N_{2}:\rangle}$ being the generalized Cauchy-Schwarz parameter
introduced in the main text. The covariance of two correlated beams obtained by splitting a single
thermal beam is $\langle \delta N_{1}\delta N_{2}\rangle_{TH}= M \eta_{1}\eta_{2} \mu_{TH}^{2}$,
while the one of TWB is $\langle \delta N_{1}\delta N_{2}\rangle_{TW}= M \eta_{1}\eta_{2} \mu_{TW}
(1+\mu_{TW})$ (see for example Ref. [27]). By using this relation with the assumption of the same
local resources, $\mu_{TH}=\mu_{TW}=\mu$ we can derive explicitly $R\approx(1+\mu)/\mu$, which is
insensitive to the amount of noise and loss. On the other side the generalized Cauchy-Schwarz
parameter for a split thermal beam is $\varepsilon_{0}^{(CI)}=1$, where the subscript "$0$" stands
for "in absence of background", as it can be easily derived from the equations of covariance and
single beam fluctuations used previously.

As described in the text $\varepsilon_{0}^{(CI)}=1$ represents the best result for classical states
and, in this sense, the comparison with split thermal beams represents the comparison with the
"best" classical case.


\begin{thebibliography}{30}

\bibitem{4} D. Bouwmeester {\it et al.}, Nature {\bf 390}, 575 (1997).
\bibitem{6} D. Boschi, S. Branca, F. De Martini, L. Hardy, and S. Popescu, Phys. Rev. Lett. {\bf 80}, 1121 (1998).
\bibitem{7} J. L. O'Brien, Science {\bf 318}, 1567 (2007).
\bibitem{8} X. Yao \textit{et al.}, Nature {\bf 482}, 489 (2012).
\bibitem{9} T. Yamamoto, M. Koashi, S. K. \"Ozdemir, and N. Imoto, Nature {\bf 421}, 343 (2003).
\bibitem{10} J. W. Pan, C. Simon, C. Brukner, and A.  Zeilinger, Nature {\bf 410}, 1067 (2001).
\bibitem{11} J. W. Pan, S. Gasparoni, R. Ursin, G. Weihs, and A. Zeilinger, Nature {\bf 417}, 4174 (2003).

\bibitem{Kolobov} M. I. Kolobov editor, \emph{Quantum Imaging}, (Springer, New York, 2007).

\bibitem{Treps:03} N. Treps {\it et al.}, Science {\bf 301}, 940 (2003).

\bibitem{boy:08} V. Boyer, A. M. Marino, R. C. Pooser, and P. D. Lett,  Science {\bf 321}, 544 (2008).

\bibitem{bri:10} G. Brida, M. Genovese, and I. Ruo Berchera, Nature Photonics {\bf 4}, 227 (2010).

\bibitem{gio:11} V. Giovannetti, S. Lloyd, and L. Maccone, Nat. Phot. {\bf 5}, 222 (2011).

\bibitem{Walmsley-PRL-2011} N. Thomas-Peter {\it et al.}, Phys. Rev. Lett. {\bf 107}, 113603 (2011).

\bibitem{llo:08} S. Lloyd, Science {\bf 321}, 1463 (2008).

\bibitem{tan:08} S. Tan {\it et al.}, Phys. Rev. Lett. {\bf 101}, 253601 (2008).

\bibitem{sha:09} J. H. Shapiro and S. Lloyd, New Journ. of Phys. {\bf 11}, 063045 (2009).

\bibitem{gua:09} S. Guha and B. I. Erkmen, Phys. Rev. A {\bf 80}, 052310 (2009).



\bibitem{ms} M. F. Sacchi, Phys. Rev. A {\bf 71}, 062340 (2005); {\bf 72}, 014305 (2005).

\bibitem{sha:PRA:09} J. Shapiro, Phys. Rev. A {\bf 80}, 022320 (2009).


\bibitem{aud:07} K. M. R. Audenaert {\it et al.},  Phys. Rev. Lett. {\bf 98}, 160501 (2007).

\bibitem{cal:08} J. Calsamiglia, R. Munoz-Tapia, Ll. Masanes, A. Acin, and E. Bagan, Phys. Rev. A {\bf 78}, 012331 (2008).

\bibitem{pir} S. Pirandola and S. Lloyd, Phys. Rev. A {\bf 77}, 032311 (2008).


\bibitem{bri:ea} G. Brida {\it et al.}, Phys. Rev. Lett. {\bf 108}, 253601 (2012)


 \bibitem{tak} H. Takahashi {\it et al.}, Phys. Rev. Lett. {\bf 101}, 233605 (2008).
 \bibitem{al} J. B. Altepeter {\it et al.}, Phys. Rev. Lett. {\bf 90}, 193601 (2003).


 \bibitem{Brambilla-PRA-2004} E. Brambilla, A. Gatti, M. Bache, and L. A. Lugiato, Phys. Rev. A \textbf{69}, 023802 (2004).



\bibitem{Bri-OptExp:10} G. Brida {\it et al.}, Opt. Exp. {\bf 18}, 20572 (2010).


\bibitem{pra} G. Brida, M. Genovese, A. Meda, and I. Ruo Berchera, Phys. Rev. A {\bf 83}, 033811 (2011).

\bibitem{Sekatski2012} P. Sekatski {\it et al}, J. Phys. B  {\bf 45}, 124016 (2012).


\bibitem{m} T. Iskhakov, M. V. Chekhova, and G. Leuchs, Phys. Rev. Lett. {\bf 102}, 183602 (2009).






\end{thebibliography}
\end{document}